\documentclass[aps,prd,showpacs,nofootinbib,10pt,superscriptaddress,twocolumn,amsmath,amssymb,floatfix,showkeys]{revtex4-1}
\usepackage{color,amsmath,amssymb,graphicx,latexsym,lineno}
\usepackage{threeparttable,multirow,txfonts,booktabs}
\usepackage{accents,slashed,ulem,subfig}
\usepackage[para]{footmisc}
\definecolor{blue0}{rgb}{0,0,0.6}
\usepackage[colorlinks,linkcolor=blue0,anchorcolor=blue0,citecolor=blue0,urlcolor=blue0]{hyperref}

\newlength{\dhatheight}
\newcommand{\doublehat}[1]{%
    \settoheight{\dhatheight}{\ensuremath{\hat{#1}}}%
    \addtolength{\dhatheight}{-0.15ex}%
    \hat{\vphantom{\rule{1pt}{\dhatheight}}%
    \smash{\hat{#1}}}}
\graphicspath{{figs/}}

\begin{document}

\title{Constraints on Axion-like Particles from the Observation of GRB 221009A by LHAASO}

\author{Lin-Qing Gao}
\affiliation{School of Nuclear Science and Technology, University of South China, Hengyang 421001, China}
\author{Xiao-Jun Bi}
\affiliation{Key Laboratory of Particle Astrophysics, Institute of High Energy Physics,
Chinese Academy of Sciences, Beijing 100049, China}
\affiliation{School of Physical Sciences, University of Chinese Academy of Sciences, Beijing, China}
\author{Jun Li}
\affiliation{Key Laboratory of Dark Matter and Space Astronomy, Purple Mountain Observatory, Chinese Academy of Sciences, 210033 Nanjing, Jiangsu, China}
\affiliation{School of Astronomy and Space Science, University of Science and Technology of China, 230026 Hefei, Anhui, China}
\author{Run-Min Yao}
\affiliation{Key Laboratory of Particle Astrophysics, Institute of High Energy Physics,
Chinese Academy of Sciences, Beijing 100049, China}
\affiliation{School of Physical Sciences, University of Chinese Academy of Sciences, Beijing, China}
\author{Peng-Fei Yin}
\affiliation{Key Laboratory of Particle Astrophysics, Institute of High Energy Physics,
Chinese Academy of Sciences, Beijing 100049, China}


\begin{abstract}

The LHAASO collaboration recently reported the measurement of the gamma-ray spectra of GRB 221009A, which is the brightest burst ever, covering an energy range from 0.3 $\mathrm{TeV}$ to about 10 $\mathrm{TeV}$. Based on the observation, we investigate the ALP-photon oscillation effect in the host galaxy of GRB 221009A and the Milky Way. The ${\rm CL_s}$ method is applied to set constraints on the ALP parameters in this study. Given the uncertain magnetic field configuration in the host galaxy, we use three different models: a homogeneous magnetic field model, a magnetic field model identical to that of the Milky Way, and a model constructed from the HST observations of the host galaxy. We find that the constraints derived using these three host galaxy magnetic field models are comparable. Our results are complementary in the small ALP mass regions compared with other experiments.

\end{abstract}

\maketitle

\section{introduction}

The light pseudoscalar particle axion was originally predicted by the model solving the strong CP problem in quantum chromodynamics \cite{Peccei:1977ur,Peccei:2006as,Weinberg:1977ma, Wilczek:1977pj}. In addition to axions, many models that beyond the Standard Model, such as string theory \cite{Svrcek:2006yi, Arvanitaki:2009fg, Marsh:2015xka}, commonly predict the existence of similar pseudoscalar particles, which are called axion-like particles (ALPs). 
Both axion and ALPs have been widely discussed due to their broad connections with several important problems in particle physics and cosmology. For example, they can be good candidates as dark matter \cite{Preskill:1982cy,Abbott:1982af, Dine:1982ah,Sikivie:2009fv}. 

The effective coupling between the ALP and photon can lead to ALP-photon conversion in external magnetic fields. This phenomenon has been extensively studied in various situations, including laboratory experiments, cosmological observations, and astrophysical observations. Given the presence of astrophysical magnetic fields at large scale, many astrophysical objects are ideal sources to study the ALP-photon oscillation effect \cite{DeAngelis:2007dqd, Hooper:2007bq, Simet:2007sa, Mirizzi:2007hr, Belikov:2010ma, DeAngelis:2011id, Horns:2012kw, HESS:2013udx, Meyer:2013pny, Tavecchio:2014yoa, Meyer:2014epa, Meyer:2014gta, Fermi-LAT:2016nkz, Meyer:2016wrm, Berenji:2016jji,
Xia:2018xbt,
Galanti:2018nvl,
Galanti:2018upl, Galanti:2018myb, Zhang:2018wpc, Liang:2018mqm, Bi:2020ths, Guo:2020kiq, Li:2020pcn,  Cheng:2020bhr, Liang:2020roo,Davies:2020uxn,Li:2021gxs,Gao:2023dvn,Pant:2023khq}. 
The ALP-photon mixing effect could potentially alter the observed spectra of photons from these objects. In the case of extragalactic sources, photons could undergo a conversion into ALPs within the ambient magnetic field of the source. These ALPs then traverse through cosmic space without impediment and eventually convert back into photons in the magnetic field of the Milky Way. These processes are expected to reduce the absorption effects of extragalactic background light (EBL) on high-energy photons and cause substantial alterations to the observed energy spectra.

GRB 221009A, a gamma-ray burst (GRB) with a redshift  of 0.151 \cite{z0.15_1,Malesani:2023aom,GCN32648_X-shooter}, has been recognized as the most luminous long GRB ever observed \cite{Burns:2023oxn, Lan:2023khy, Frederiks:2023bxg}. Initially reported by Fermi-GBM \cite{GRB221009A_GBM}, it was subsequently observed by multiple detectors across a wide energy band in 2022, ranging from radio to very high energy (VHE) $\gamma$-rays \cite{GRB221009_X,GRB221009_X2,GRB221009_X3,GRB221009_near-infrared,GRB221009_ATA,GRB221009_optical1,GRB221009_REM,GCN32637_LAT,GCN32661_STIX,LHAASO_GRB221009,Williams:2023sfk,LHAASO:2023kyg}. Among these observations, the Large High Altitude Air Shower Observatory (LHAASO) reported remarkably striking results, detecting more than 5000 VHE photons with the highest energy photon estimated to be $\sim18$ $\mathrm{TeV}$ \cite{LHAASO_GRB221009}. The detection of such high-energy photons from extragalactic sources poses a challenge to many EBL models, as it is difficult for high-energy photons with energies of $\sim 18$ $\mathrm{TeV}$ to reach the Earth according to these models. The exceptional nature of this high-energy observation has attracted considerable attention, leading to various proposed explanations, including Lorentz invariance violation and ALPs \cite{Galanti:2022pbg,Li:2022wxc,Baktash:2022gnf,Troitsky:2022xso, Gonzalez:2022opy,Galanti:2022xok,Zhang:2022zbm, Carenza:2022kjt,Vardanyan:2022ujc}.

Recently, the results of the spectral energy distribution of GRB 221009A observed by LHAASO have been published \cite{Cao:2023til}. The energy range spans from 0.3 $\mathrm{TeV}$ to about 10 $\mathrm{TeV}$, encompassing observations from the WCDA and KM2A detectors, thus establishing it as the highest energy GRB spectrum up to date. 
The highest energy event is reconstructed to be  $12.2^{+3.5}_{-2.4}$ $\mathrm{TeV}$ through a comprehensive analysis, 
assuming that the spectrum follows a power-law with an exponential cut-off function.
The observed spectra do not necessitate the introduction of new physics, such as ALPs. Consequently, the collaboration has derived constraints on ALPs based on these findings \cite{Cao:2023til}. 

The ALP-photon oscillation effect depends on the astrophysical magnetic fields, and various studies have employed diverse models to characterize these fields. As a simple approximation, in Ref.~\cite{Cao:2023til,Baktash:2022gnf} the magnetic field in the host galaxy (HG) of GRB 221009A is assumed to be a uniform magnetic field of 0.5 $\mu \rm G$ with a length of $10~\mathrm{kpc}$. To better reflect the actual galactic environment, Ref.~\cite{Gonzalez:2022opy, Carenza:2022kjt} assume that the magnetic field configuration of the HG is identical to that of the Milky Way. Given the limited available information about the HG,  Ref.~\cite{Carenza:2022kjt}  investigates the influence of photon trajectories in the magnetic field.
Fortunately, recent observations of Hubble Space Telescope (HST) have revealed important properties of the HG \cite{Levan:2023doz}.
These observations indicate that the HG has a disc-like structure and is observed close to edge-on. Furthermore, GRB 221009A is found to be close to the core regions of the HG with an offset of $0.25''$. Based on these findings, an improved HG magnetic field model is constructed in Ref.~\cite{Troitsky:2023uwu}. Additionally, a detailed analysis of the HG magnetic field is available in Ref.~\cite{Galanti:2022pbg}.

In this study, we focus on examining the influence of the HG magnetic field of GRB 221009A on the ALP effect, deriving constraints from the observation by LHAASO. Three magnetic field models are considered, including the simple homogeneous magnetic field, the  Jansson \& Farrar magnetic field model of the Milky Way \cite{Jansson:2012rt}, and the improved magnetic field model proposed by Ref.~\cite{Troitsky:2023uwu}. We establish constraints on the ALP parameters using the ${\rm CL_s}$ method \cite{Read:2002hq_cls} as \cite{Gao:2023dvn}, and compare these constraints derived from different magnetic field models. 

This paper is organized as follows. In section \ref{sec:ALP-photon}, we briefly introduce the ALP-photon oscillation effect and the HG magnetic fields of GRB 221009A. In section \ref{sec:method}, we describe the spectrum fitting method and the ${\rm CL_s}$ method that we use to set constraints on the ALP parameters. In section \ref{sec:results}, we present the constraints on the ALP parameters obtained from the observations of GRB 221009A by LHAASO. Finally, we conclude in section \ref{sec:conclusion}.

\section{ALP-photon oscillation and Astrophysical environments}\label{sec:ALP-photon}

In this section, we briefly outline the ALP-photon oscillation effect in astrophysical magnetic fields. We then introduce three HG magnetic field models for GRB 221009A. Finally, we provide a brief overview of the effects of the galactic magnetic field and EBL attenuation.

The state of the photon together with ALP in propagation can be described by the polarization density matrix, denoted as $\rho \equiv \Psi \otimes \Psi^\dagger$, where $\Psi = (A_1, A_2, a)^T$. Here, $a$ represents the ALP, and $A_1$ and $A_2$ represent the two photon transverse polarization amplitude along the directions of $\boldsymbol{x_1}$ and $\boldsymbol{x_2}$, respectively. When the ALP-photon beam propagates along the $\boldsymbol{x_3}$ direction in a homogeneous magnetic field, the density matrix follows a von Neumann-like commutator equation:
\begin{equation}\label{equ:von Neumannn-like}
i\frac{d\rho}{d\boldsymbol{x_3}} = [\rho, \mathcal{M}_0],
\end{equation}
where $\mathcal{M}_0$ is the mixing matrix that encompasses the effects of the environment and the interaction between photons and ALPs. Detailed expressions and solutions for this equation can be found in the literature, such as Ref. ~\cite{DeAngelis:2011id, Galanti:2018nvl, Davies:2020uxn}.

Astrophysical magnetic fields are present along the line of sight. The path of photons traveling from the source to Earth can be divided into numerous segments, with the magnetic field considered constant within each segment. By solving Eq. \ref{equ:von Neumannn-like}, the final survival probability of the photons can be expressed as 
\begin{equation}\label{equ:P_ga}
P_{\gamma\gamma} = \mathrm{Tr}\left((\rho_{11}+\rho_{22})\mathcal{T}(\boldsymbol{x_3}) \rho(0) \mathcal{T}^{\dagger}(\boldsymbol{x_3})\right), \end{equation}
where $\mathcal{T}(\boldsymbol{x_3}) = \prod \limits_i^n \mathcal{T}_i (\boldsymbol{x_3})$ represents the product of transfer matrices derived from the von Neumann-like commutator equation for each segment, $\rho(0) = \mathrm{diag}(1/2, 1/2, 0)$ is the density matrix for the initially unpolarized photon beam emitted from the source, $\rho_{11}=\mathrm{diag}(1, 0, 0)$, and $\rho_{11}=\mathrm{diag}(0, 1, 0)$.

Photons from GRB 221009A traverse various astrophysical environments, including the HG, extragalactic space, and the Milky Way. A significant theoretical uncertainty arises from the limited knowledge about the properties of the HG.
Recent observations from the HST have revealed that GRB 221009A is located in a peculiar HG, displaying characteristics of an edge-on system \cite{Levan:2023doz}. This HG is a representative star-forming galaxy with an effective radius of $2.45 \pm 0.20~\mathrm{kpc}$ and a stellar mass of ${\rm log}(M/M_{\odot})=9.00^{-0.47}_{+0.23}$, where $M_{\odot}$ denotes the mass of the Sun. GRB 221009A is located in proximity to the central region of the HG, with an offset of $0.25''$, i.e. approximately $0.65 ~\mathrm{kpc}$. The distance of GRB to the disk plane is approximately $0.44~\mathrm{kpc}$, and the projected parallel distance to the HG center is approximately $0.48~\mathrm{kpc}$ ~\cite{Troitsky:2023uwu}. 

The uncertain nature of the magnetic field within the HG induces some uncertainties in the calculated survival probability of photons, consequently impacting the constraints on the ALP parameters. To enhance the reliability of the constraints, we explore and compare three distinct HG magnetic field models for GRB 221009A:
\begin{itemize} 
\item Model 1: a homogeneous magnetic field throughout the HG. 
\item Model 2: a magnetic field model of the Milky Way. 
\item Model 3: a model constructed from the HST observations \cite{Levan:2023doz}, combined with the measurements and simulations of magnetic field for other galaxies  \cite{Troitsky:2023uwu}. 
\end{itemize}

\paragraph*{Model 1}Observations indicate that numerous galaxies exhibit coherent magnetic fields across kiloparsec scales, typically with magnitudes of the order of $\mu\mathrm{G}$. The authors of Ref. \cite{Cao:2023til,Baktash:2022gnf} adopt Model 1 as a minimum scenario, which assumes a 0.5 $\mu\mathrm{G}$ homogeneous transverse magnetic field component with a length of $10~\mathrm{kpc}$. In this study, we take the homogeneous magnetic field strength to be 1 $\mu\mathrm{G}$ as a typical value.  

\paragraph*{Model 2}The magnetic field of the Milky Way is an extensively studied phenomenon. Detailed models have been developed based on many measurements, such as Faraday rotation measures and polarized synchrotron emission. 
In studies like Ref.~\cite{Gonzalez:2022opy,Carenza:2022kjt}, Model 2 is adopted, assuming that the magnetic field within the HG under investigation is identical to that of the Milky Way.

\paragraph*{Model 3}The author of Ref. \cite{Troitsky:2023uwu} presents an improved magnetic field model for the HG based on the HST observations. The construction of this model involves three steps.
In the first step, a smoothing technique is applied to refine a  magnetic field model of the Milky Way in the central region. This model comprises a disk component and a toroidal halo component. In the second step, an additional X-shape component is introduced, and modifications are made to match the observations of NGC 891, which is a nearby edge-on disk galaxy. The choice of NGC 891 is motivated by its compactness and higher star formation rate, which make its properties more similar to those expected for the HG under investigation. In the final step, the author scales the field strength and adjusts some parameters to account for the differences between NGC 891 and the HG. Detailed expressions for each component of the HG magnetic field can be found in the appendix of Ref. \cite{Troitsky:2023uwu}.

Note that despite the existing HG magnetic field models, there remain some unknown parameters related to the position of the GRB in the HG, which impact the paths of photons\cite{Troitsky:2023uwu}. Two crucial uncertainties are highlighted here. Firstly, while the projected distance from the GRB to the HG center can be derived from the HST observations, the precise location of GRB 221009A within the HG along the line of sight remains ambiguous. To account for this, the position of GRB 221009A along the line of sight is characterized by a free parameter $y_0$. Here, $y_0$ denotes the distance from the GRB to the plane $\mathcal{P}$ containing the HG center, which is perpendicular to the line of sight. 
Secondly, given models characterized by a polar angle $\theta$, the  orientation of the polar axis in the disk plane of the HG is uncertain. We define $\theta_0$ as the angle between the polar axis and the plane $\mathcal{P}$. This parameter $\theta_0$ is also treated as a free parameter in Model 2 and Model 3. These free parameters are essential for accurately determining the position of the GRB and for ascertaining the specific photon  trajectory within the HG.

In this study, we do not consider the ALP-photon oscillation in the GRB jet or extragalactic space, as their contributions in these regions are likely to be negligible.  Comprehensive discussions on these contributions can be found in Ref.~\cite{Galanti:2022pbg,Galanti:2022xok}.
The oscillation effect in the Milky Way is taken into account here. The magnetic field of the Milky Way can be characterized by a regular component with a small-scale turbulent component. In this study, we focus on the regular component and neglect the turbulent component due to its limited coherent length.
For the regular component of the galactic magnetic field, we adopt the model outlined in Ref.~\cite{Jansson:2012rt}.
Additionally, we use the NE2001 model to characterize the electron density distribution of the Milky Way \cite{Cordes:2002wz} and extend its application to the HG. 

At high energies, photons emitted by the source may undergo absorption by the EBL through the pair production process $\gamma + \gamma_{EBL} \to e^+ + e^-$. This absorption results in a reduction in the photon flux, quantified by a factor of $e^{-\tau_{\gamma}}$, where $\tau_\gamma$ denotes the optical depth. The value of $\tau_\gamma$ relies on the redshift of the source and the distribution of the EBL. In this study, we adopt the EBL model detailed in Ref. \cite{Saldana-Lopez:2020qzx} and obtain the optical depth using the \textit{ebltable} package \cite{manuel_meyer_2022_7312062}. This EBL model can well explain the LHAASO observations \cite{Cao:2023til}.

After taking into account the effects described above, the observed photon spectrum on Earth is given by 
\begin{equation}
\label{equ:df/dE} \frac{d\Phi}{dE} = P_{\gamma\gamma}\frac{d\Phi_{int}}{dE}, 
\end{equation} 
where $P_{\gamma\gamma}$ is the survival probability of photons after propagation, and $d\Phi_{int}/dE$ is the intrinsic spectrum.
The survival probability $P_{\gamma\gamma}$ is numerically calculated from Eq.~\ref{equ:von Neumannn-like} and Eq.~\ref{equ:P_ga}. 
The transfer matrix $\mathcal{T}$ mainly consists of the contributions in three regions: the ALP-photon oscillation effect in the HG and Milky Way, as well as the absorption effect induced by EBL in extragalactic space. If ALP is not considered, the survival probability is simply given by $P_{\gamma\gamma} = e^{-\tau_\gamma}$.

\section{method}\label{sec:method}

In this section, we outline the approach used to fit the GRB spectra and the statistical method employed to establish constraints on the ALP parameters. GRB 221009A was initially detected by Fermi-GBM at 13:16:59.99 UT of the October 9th, 2022, denoted as $T_0$. The data from the LHAASO observation of GRB 221009A are divided into two time intervals based on the light curve observed by KM2A. The first interval spans from $T_0$+230 s to $T_0$+300 s; the second interval covers $T_0$+300 s to $T_0$+900 s. Detailed spectra, encompassing observations from both KM2A and WCDA, for each interval are provided in \cite{Cao:2023til}.

For the intrinsic spectrum of the GRB, we adopt a log-parabolic function $d\Phi/dE_{int} = F_0(E/E_0)^{-\Gamma - b\mathrm{log} (E/E_0)}$, where $F_0$, $\Gamma$, and $b$ are three free parameters, and $E_0$ is set to be 1 $\mathrm{TeV}$. Considering the effects of ALP-photon oscillation and EBL attenuation, we obtain the predicted photon spectrum with Eq.~\ref{equ:df/dE}. 

Given the specific values of the ALP parameters $m_a$ and $g_{a\gamma}$, where $m_a$ and $g_{a\gamma}$ represent the mass of ALP and its coupling strength with photons, respectively, the best-fit values of the spectrum are obtained by minimizing the $\chi^2$ function
\begin{equation}
    \chi^2(F_0, \Gamma, b; m_a, g_{a\gamma}) = \sum\limits _{i=1}^{N} \frac{\left(\frac{d\Phi}{dE}(F_0, \Gamma, b; m_a, g_{a\gamma})|_i - \frac{d\Phi}{dE} |_{\rm{obs}, i}\right)^2}{\delta _i^2},
\end{equation}
where $i$ represents the $i$-th energy bin, $\frac{d\Phi}{dE}$, $\frac{d\Phi}{dE} |_{\rm obs}$, and $\delta$ denote the predicted value, observed value, and experimental uncertainty of the photon flux, respectively.  
In order to mitigate the oscillation effect that may not be present in the observation due to the limited experimental resolution, the predicted flux in each bin has been averaged.
Subsequently, the likelihood function is given by
\begin{equation}
\label{equ:likelihood}
    \mathcal{L} (F_0, \Gamma, b; m_a, g_{a\gamma}) = \exp(-\chi^2(F_0, \Gamma, b; m_a, g_{a\gamma})/2).
\end{equation}

In order to set constraints, we define the test statistic (TS) value for a given data sample as the logarithmic likelihood ratio
\begin{equation}\label{equ:TS_CLs}
    TS(m_a, g_{a\gamma}) = -2\ln\left( \frac{\mathcal{L}_1 (\doublehat{F_0}, \doublehat{\Gamma}, \doublehat{b}; m_a, g_{a\gamma})}{\mathcal{L}_0(\hat{F_0}, \hat{\Gamma}, \hat{b})} \right),
\end{equation}
where $\mathcal{L}_0$ represents the maximum value of the likelihood function under the null hypothesis without the ALP-photon oscillation effect, and $\mathcal{L}_1$ represents the maximum value of the likelihood under the alternative hypothesis, including ALP-photon oscillation effect with the two parameters $m_a$ and $g_{a\gamma}$. Here, $(\hat{F_0}, \hat{\Gamma}, \hat{b})$ and $(\doublehat{F_0}, \doublehat{\Gamma}, \doublehat{b})$ denote the best-fit values of the parameters of the intrinsic spectrum, under the null and alternative hypotheses, respectively.
Due to the non-linear impact of the ALP parameters on the photon spectrum, Wilks' theorem \cite{wilk} is not applicable in this scenario \cite{Fermi-LAT:2016nkz}. Consequently, the distribution of TS value can not be well described by a $\chi^2$ distribution. Therefore, in order to derive constraints on the ALP parameters in this case, Monte Carlo simulations are necessary.

In this study, we employ the ${\rm CL_s}$ method \cite{Read:2002hq_cls}, which is widely used in high energy experiments, to set constraints on the ALP parameters, following the same procedure described in Ref.~\cite{Gao:2023dvn}. Within the $(m_a, g_{a\gamma})$ parameter space, we perform a scan of numerous parameter points. For each parameter point, we evaluate whether it has been excluded by the LHAASO observation.

We choose a parameter point within the $(m_a, g_{a\gamma})$ parameter space as an illustrative example. At this chosen parameter point, a predicted photon spectrum is obtained by maximizing the likelihood function Eq.~\ref{equ:likelihood} for the actual data. Subsequently, based on this predicted spectrum, we generate a mock data set comprising $10^4$ samples. For a mock data sample, the central value of the photon flux in each energy bin is randomly generated according to a Gaussian distribution, with the mean value and deviation set to be the predicted flux and the experimental uncertainty, respectively; the uncertainty of the photon flux is also taken to be the experimental value.
We then calculate the TS value for each mock data sample using Eq.~\ref{equ:TS_CLs}, and derive a TS distribution from all the samples, denoted as $\rm \{TS\}_{s+b}$. Using the same method, we can also derive the TS distribution $\rm \{TS\}_{b}$ based on another mock data set, which is generated with the predicted photon spectrum assuming no ALPs.

Subsequently, we calculate the TS value $\rm TS_{obs}$ for the actual observed data. According to $\rm \{TS\}_{s+b}$, the probability of finding a TS value larger than $\rm TS_{obs}$ can be calculated, denoted as $\rm CL_{s+b}$. In traditional statistical analysis, this probability serves as a criterion for determining whether the parameter point is excluded by the observation. For instance, if $\rm CL_{s+b}$ is less than 0.05, the parameter point is deemed to be excluded at a 95\% confidence level (C.L.). However, in certain cases where it is challenging to distinguish between $\rm \{TS\}_{s+b}$ and $\rm \{TS\}_{b}$
, this criterion may not yield a definitive conclusion.
In order to address the impact of $\rm \{TS\}_{b}$, ${\rm CL_s}$ is defined as
\begin{equation}
\rm CL_s=\frac{CL_{s+b}}{CL_{b}},
\end{equation}
where $\rm CL_b$ denotes the probability of obtaining a TS value greater than $\rm TS_{obs}$ based on $\rm \{TS\}_{b}$. The $\rm CL_s$ method utilizes the $\rm CL_s$ value as a criterion to determine whether this parameter point is excluded. For instance, if the $\rm CL_s$ value of the  parameter point is smaller than 0.05, this point is considered to be excluded at a 95\% C.L..

\begin{figure*}
  \centering
  \subfloat[$m_a$ = $10^{-10}$ eV, $g_{a\gamma}$ = $1.6\times10^{-11}$ GeV.]
{\includegraphics[width=0.4\textwidth]{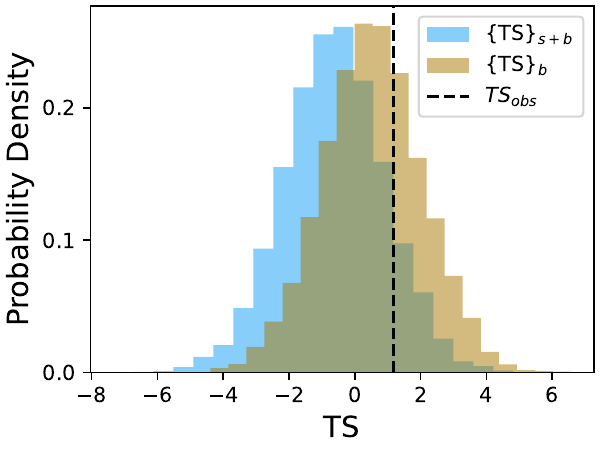}\label{fig: TS distri 1}}
  \subfloat[$m_a$ = $10^{-8}$ eV, $g_{a\gamma}$ = $5.0\times10^{-11}$ GeV.]
{\includegraphics[width=0.4\textwidth]{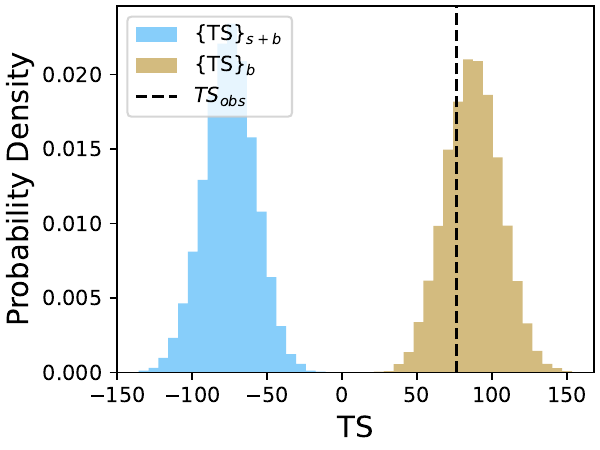}\label{fig: TS distri 2}} 
  \caption{The TS distributions $\rm \{TS\}_{b}$ and $\rm \{TS\}_{s+b}$ for two parameter points $(m_a, g_{a\gamma})= (10^{-10} \rm eV, 1.6\times 10^{-11} \rm GeV^{-1})$ and $(10^{-8} \rm eV, 5.0\times 10^{-11} \rm GeV^{-1})$ are shown in panel (a) and (b), respectively. The vertical dashed lines represent the observed TS value. Here we take the HG magnetic field model as Model 3, with ($\theta_0$, $y_0$) = ($0^\circ$, 0 kpc).}
  \label{fig: TS distri}
\end{figure*}

In FIG.~\ref{fig: TS distri}, we show the distributions $\rm \{TS\}_{b}$ and $\rm \{TS\}_{s+b}$ for two parameter points $(m_a, g_{a\gamma})= (10^{-10} \rm eV, 1.6\times 10^{-11} \rm GeV^{-1})$ and $(10^{-8} \rm eV, 5.0\times 10^{-11} \rm GeV^{-1})$ utilizing the HG magnetic field 
Model 3 with ($\theta_0$, $y_0$) = ($0^\circ$, 0), denoted as points A and B, respectively. The corresponding $\rm CL_s$ values for points A and B are 0.34 and 0.0, respectively. These results indicate that point B can be excluded at the 95\% C.L., while the point A is not excluded at the 95\% C.L.. Note that the criteria of $\rm CL_s$ and $\rm CL_{s+b}$ yield congruent conclusions for parameter points similar to point B, where the two TS distributions $\rm \{TS\}_{b}$ and $\rm \{TS\}_{s+b}$ are distinctly separated. For certain parameter points similar to point A, characterized by a substantial overlap between $\rm \{TS\}_{b}$ and $\rm \{TS\}_{s+b}$, it is possible that they are excluded by the $\rm CL_{s+b}$ criterion, but remain allowed by the $\rm CL_{s}$ criterion at a same C.L., owing to $\rm CL_{s} \gg CL_{s+b}$ in this case.

Following the outlined method, we can apply it to all the parameter points within the parameter space and determine whether they are excluded at a 95\% C.L. Based on the results obtained, we can establish the excluded regions in the parameter space.

\section{results}\label{sec:results}

In this section, we present the constraints on the ALP parameters using the spectra of GRB 221009A observed by LHAASO. Note that under the null hypothesis, the best-fit reduced $\chi^2$ values for the two time intervals mentioned above are 11.3/9 and 6.6/10, respectively. This indicates that the null hypothesis can well fit the observations.

Three HG magnetic field models are considered in our analysis. To address the uncertainty associated with the parameters $\theta_0$ and $y_0$, which are linked to the ambiguous position of the GRB along the line of sight, we establish constraints with different selections for these parameters. Given the HG's half-light radius of $\sim 2.45~\mathrm{kpc}$ \cite{Levan:2023doz}, we take the typical values of $y_0$ as $0~\mathrm{kpc}$, $2.5~\mathrm{kpc}$, and $-2.5~\mathrm{kpc}$. Here, $y_0<0$ indicates that the GRB is further away from us compared to $y_0>0$. For Model 1, the case $y_0=0~\mathrm{kpc}$ represents that the length of the photon path in the HG magnetic field is $10~\mathrm{kpc}$. For Model 2 and Model 3, we consider three typical values of $\theta_0$ to be $0^\circ$, $40^\circ$, and $120^\circ$.

\begin{figure*}[htbp]
  \centering
  \begin{minipage}[b]{0.4\textwidth}
    \centering
    \includegraphics[width=\textwidth]{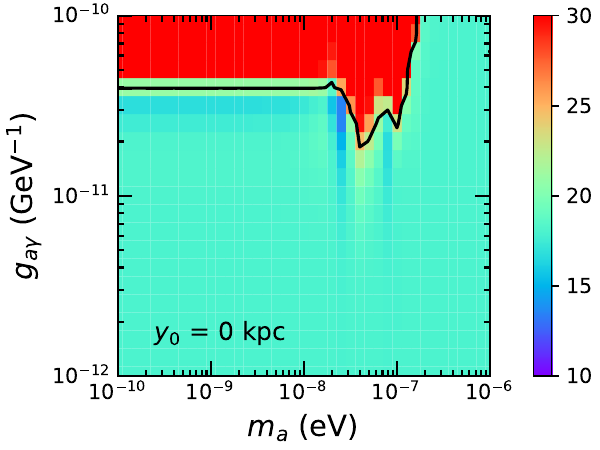}
    \caption*{(a)}
  \end{minipage}
  \begin{minipage}[b]{0.4\textwidth}
    \centering
\includegraphics[width=\textwidth]{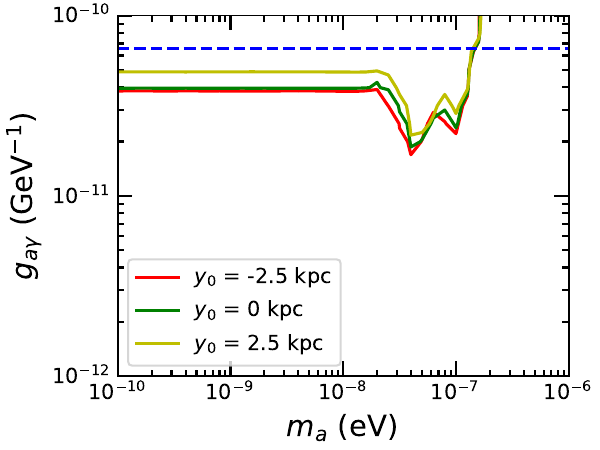}
    \caption*{(b)}
  \end{minipage}
  \caption{(a) The heat maps  illustrating the best-fit $\chi^2$ values in the $m_a-g_{a\gamma}$ plane derived with the constant HG magnetic model (Model 1). The solid black line represents the 95\% C.L. constraints from GRB 221009A obtained through the application of the ${\rm CL_s}$ method. (b) Three solid lines represent the 95\% C.L. constraints from GRB 221009A with different $y_0$. The blue dashed line represents the constraint from CAST \cite{CAST:2017uph}.}
  \label{fig: const contours}
\end{figure*}

\begin{figure*}[htbp]
  \centering
  \begin{minipage}[b]{0.4\textwidth}
    \centering
    \includegraphics[width=\textwidth]{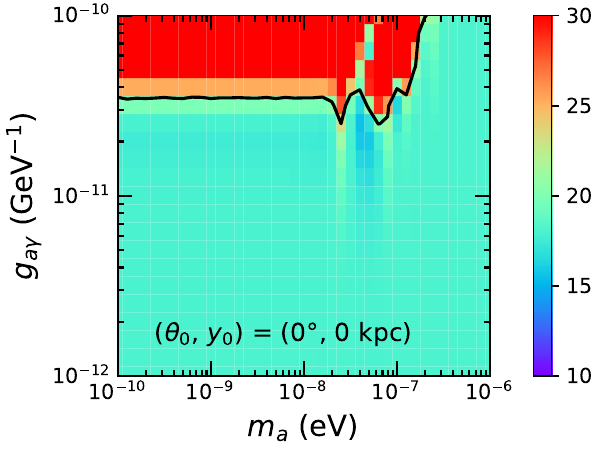}
    \caption*{(a)}
  \end{minipage}
  \begin{minipage}[b]{0.4\textwidth}
    \centering
\includegraphics[width=\textwidth]{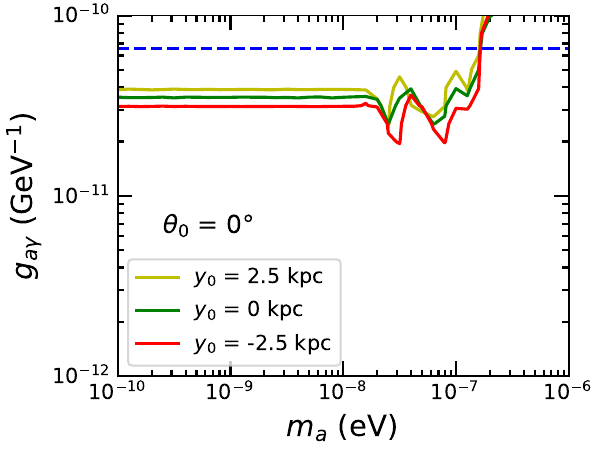}
\caption*{(b)}
  \end{minipage}
    \begin{minipage}[b]{0.4\textwidth}
    \centering
\includegraphics[width=\textwidth]{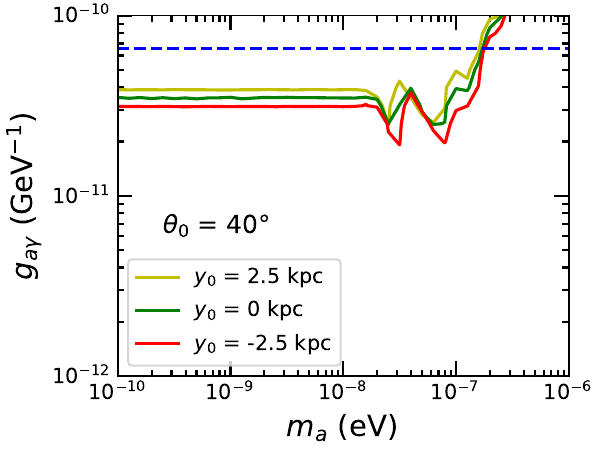}
\caption*{(c)}
  \end{minipage}
    \begin{minipage}[b]{0.4\textwidth}
    \centering
\includegraphics[width=\textwidth]{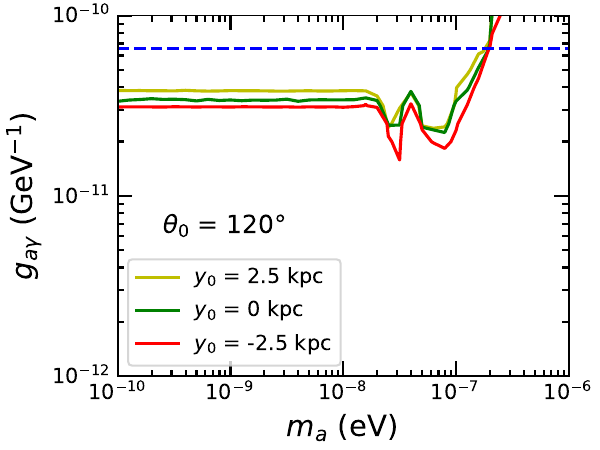}
\caption*{(d)}
  \end{minipage}
  \caption{Same as FIG. \ref{fig: const contours}, but taking the Galactic magnetic model (Model 2) as the HG magnetic model.}
  \label{fig: GMF_like contour}
\end{figure*}

\begin{figure*}[htbp]
  \centering
  \begin{minipage}[b]{0.4\textwidth}
    \centering
    \includegraphics[width=\textwidth]{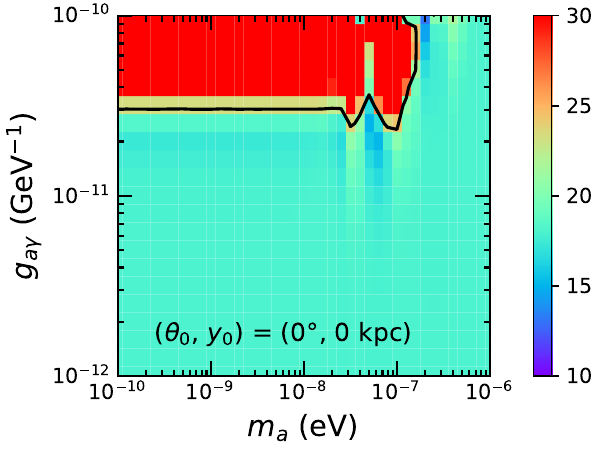}
    \caption*{(a)}
  \end{minipage}
  \begin{minipage}[b]{0.4\textwidth}
    \centering
\includegraphics[width=\textwidth]{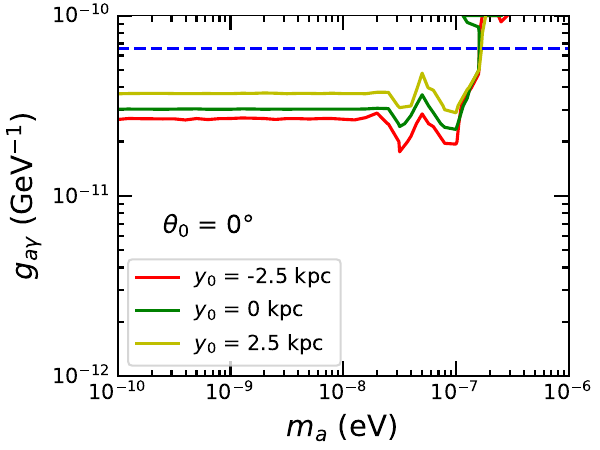}
\caption*{(b)}
  \end{minipage}
    \begin{minipage}[b]{0.4\textwidth}
    \centering
\includegraphics[width=\textwidth]{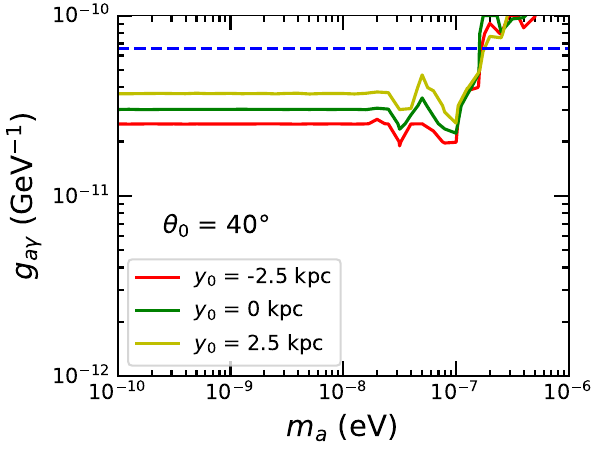}
\caption*{(c)}
  \end{minipage}
    \begin{minipage}[b]{0.4\textwidth}
    \centering
\includegraphics[width=\textwidth]{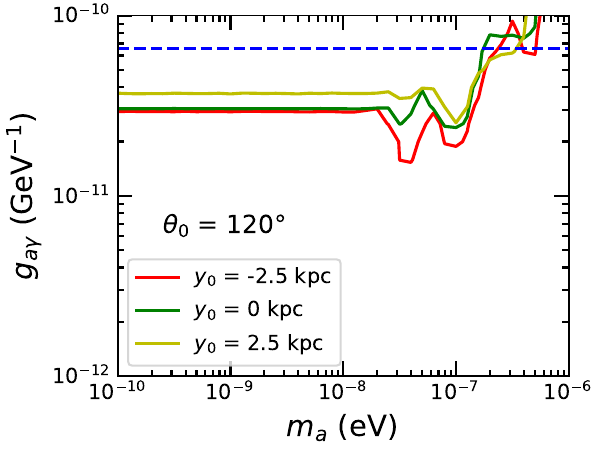}
\caption*{(d)}
  \end{minipage}
  \caption{Same as FIG. \ref{fig: const contours}, but using the HG magnetic model provided in Ref. \cite{Troitsky:2023uwu} (Model 3).}
  \label{fig: Troitsky contour}
\end{figure*}

We focus on the parameter space with $m_a \in [10^{-10},10^{-6}]$ eV and $g_{a\gamma} \in [10^{-12},10^{-10}]~\mathrm{GeV}^{-1}$, and study the combined results from two time intervals. The $\chi^2$ contours for Model 1, Model 2, and Model 3 with $\theta_0 = 0^\circ$ and $y_0$ = 0 are shown in FIG. \ref{fig: const contours} (a), FIG. \ref{fig: GMF_like contour} (a) and FIG. \ref{fig: Troitsky contour} (a), respectively. The black curve in each figure represents the constraints from GRB 221009A at 95\% C.L. with the ${\rm CL_s}$ method. In the other panels of FIG. \ref{fig: const contours}, FIG. \ref{fig: GMF_like contour}, and FIG. \ref{fig: Troitsky contour}, constraints in other cases of $\theta_0$ and $y_0$ are also presented. In these figures, the CAST constraint of $g_{a\gamma}< 6.6\times 10^{-11} \rm GeV^{-1}$ \cite{CAST:2017uph} is presented in the blue dashed line.

We find that the constraints derived with three HG magnetic field models are comparable. 
Notably, a characteristic of these constraints is that when the ALP mass is small, the constraint only depends on the coupling strength $g_{a\gamma}$. In this mass region, the mixing term $g_{a\gamma}B$ is much larger than the difference between the ALP mass term $m_a^2/E_\gamma$ and the photon effective mass term, considering the large photon energies $E_\gamma \sim \mathcal{O}\rm(TeV)$. Thus, the mixing effect between the ALP and photon almost reaches its maximum and does not depend on the ALP mass. Consequently, the photon survival probabilities are $P_{\gamma\to a}\sim \frac{1}{4}g_{a\gamma}^2B^2d^2$ with distance $d$, and the constraints are independent of the ALP mass $m_a$ in this mass region.

Upon analyzing FIG. \ref{fig: const contours}, FIG. \ref{fig: GMF_like contour}, and FIG. \ref{fig: Troitsky contour}, it becomes apparent that the constraints for $y_0=-2.5~\mathrm{kpc}$ are more stringent compared to $y_0=0~\mathrm{kpc}$ and $y_0=2.5~\mathrm{kpc}$ for the same $\theta_0$. The reason for the stronger influence of $y_0$ is that, in the case of $y_0<0$, the photon-ALP oscillation occurs over a greater distance $d$ within the HG compared to $y_0 \geq 0$. We also find that varying values of $\theta_0$ do not significantly impact the results when $y_0$ is fixed. Although the parameters $y_0$ and $\theta_0$ do have some influences on the constraints, their impacts are not substantial.

Our constraints become more stringent for ALP masses on the order of $\mathcal{O}(10^{-8}) \mathrm{eV}$ compared to smaller ALP masses. This is attributed to the difference between the ALP mass term $m_a^2/E_\gamma$ and the photon effective mass term becoming comparable to the ALP-photon mixing term $ g_{a\gamma}B$ within this parameter region, leading to subtle yet noticeable variations in the spectra. 
In our analysis, the fitting to the given observed spectra at $E_\gamma \sim \mathcal{O}\rm(TeV)$ would change sensitively due to these variations and provide more stringent constraints. However, in the data analysis of LHAASO, the constraints are set by the fitting to the observed photon events rather than the fixed energy spectrum. The variation of the constraints in the mass region $\mathcal{O}(10^{-8}) \mathrm{eV}$ does not show in Ref.~\cite{Cao:2023til} due to the different analysis approaches.

\begin{figure*}[htbp]
  \centering
    \begin{minipage}[b]{0.4\textwidth}
    \centering
    \includegraphics[width=\textwidth]{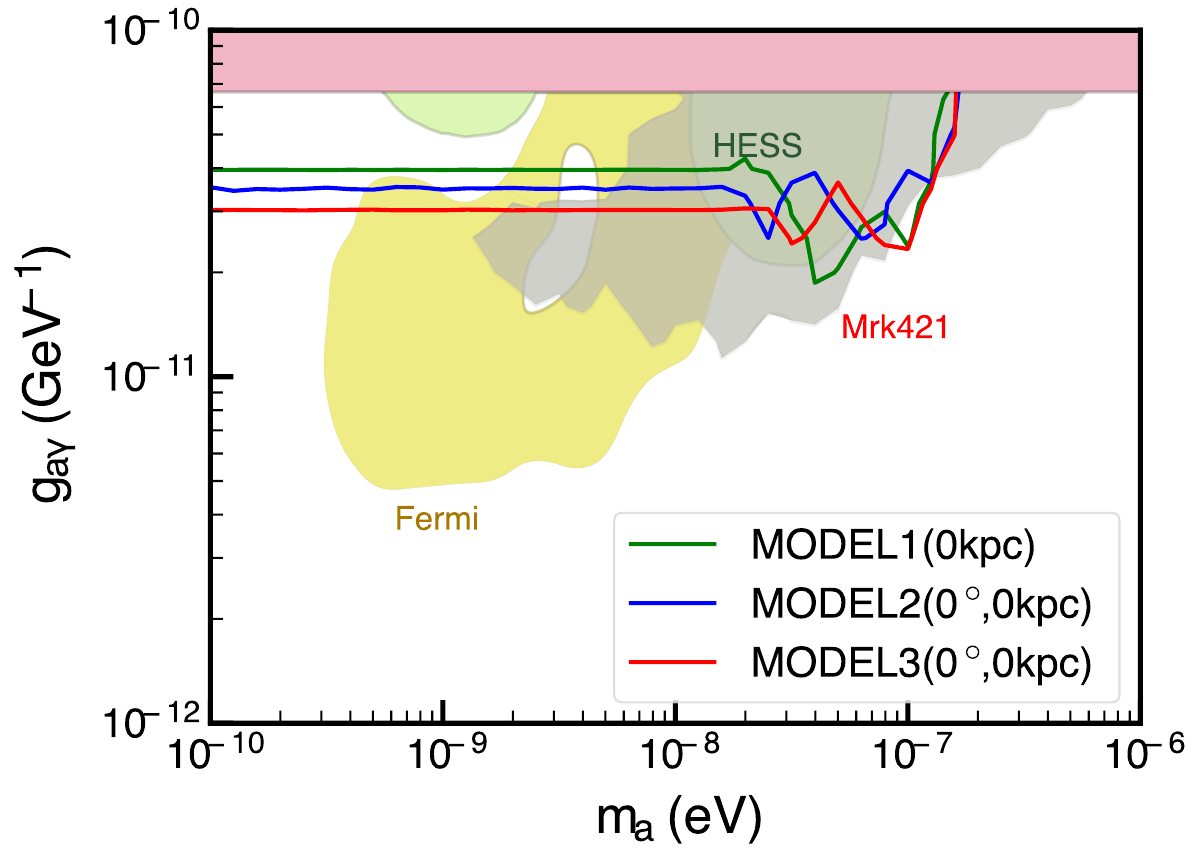}
  \end{minipage}
  \caption{The 95\% C.L. constraints on the ALP parameters obtained in this analysis. The solid lines represent the constraints derived with three HG magnetic field models with $y_0$ = 0 kpc and $\theta_0=0^\circ$. For comparison, we also display the constraints established by the CAST experiment \cite{CAST:2017uph}, the H.E.S.S. observations of PKS 2155-304 \cite{HESS:2013udx}, the Fermi-LAT observation of NGC 1275 \cite{Fermi-LAT:2016nkz}, and the MAGIC observation of Mrk 421 \cite{Gao:2023dvn} (see also \cite{AxionLimits}).
  }
  \label{fig: comp}
\end{figure*}

In FIG. \ref{fig: comp}, we present a comparison of the constraints from GRB 221009A with those from other experiments. For Model 2 and Model 3, we only include the results obtained from $(\theta_0, y_0) = (0^\circ, 0~\mathrm{kpc})$. The other experiments considered in the comparison include the CAST experiment \cite{CAST:2017uph}, Fermi-LAT observation for NGC 1275 \cite{Fermi-LAT:2016nkz}, and H.E.S.S. observation for PKS 2155-304 \cite{HESS:2013udx}. Our constraints complement the results obtained from other experiments in specific parameter regions, particularly in the small mass region.

\section{conclusion}\label{sec:conclusion}
GRB 221009A is recognized  as the most luminous long GRB observed to date. The LHAASO experiment has provided a comprehensive energy spectrum of GRB 221009A, spanning from  $0.3~\mathrm{TeV}$ to approximately $10~\mathrm{TeV}$. In this study, we set constraints on the ALP parameters using these observations. 

Based on the HST observations, it is known that the disk-like HG of GRB 221009A is viewed close to
edge-on, and GRB 221009A is in proximity to the central region of the HG. We investigate the ALP-photon oscillation phenomenon in the magnetic fields of the HG and Milky Way. Since the HG magnetic field model remains uncertain, we consider three distinct models, including a simple homogeneous magnetic field model, a magnetic field model of the Milky Way, and a model based on the HST observations. In these models, some parameters related to the position of GRB 221009A within the HG remain unknown. To address this issue, we set several constraints corresponding to various permutations of these ambiguous parameters.

The ${\rm CL_s}$ method is employed to derive 95\% C.L. constraints on the ALP parameters. Notably, the constraints obtained using the three distinct HG magnetic field models are comparable. Although the uncertain parameters $y_0$ and $\theta_0$ do exert some influences on the results, their impacts are found to be relatively minor.
The constraints derived in this study complement those from other experiments in small mass regions.

\acknowledgements
This work is supported by the National Natural Science Foundation of China under grant No. 12175248.

\newpage
\bibliographystyle{apsrev}
\bibliography{Ref}

\end{document}